\documentclass[11pt,letterpaper,notitlepage]{article}

\usepackage{amsmath,amssymb,amsfonts,graphicx,geometry,natbib,listings,url}

\setcitestyle{numbers,super,comma}

\begin{document}

\title{\bf \Large Visualizing the logistic map with a microcontroller}
\author{Juan D. Serna\thanks{serna@uamont.edu} \\
        School of Mathematical and Natural Sciences \\
        University of Arkansas at Monticello, Monticello, AR 71656 \\
\and
        Amitabh Joshi\thanks{ajoshi@eiu.edu} \\
        Department of Physics \\
        Eastern Illinois University, Charleston, IL 61920}
\date{December 25, 2011}

\maketitle

\begin{abstract}
The logistic map is one of the simplest nonlinear dynamical systems that clearly
exhibit the route to chaos. In this paper, we explored the evolution of the
logistic map using an open-source microcontroller connected to an array of light
emitting diodes (LEDs). We divided the one-dimensional interval $[0,1]$ into ten
equal parts, and associated and LED to each segment. Every time an iteration
took place a corresponding LED turned on indicating the value returned by the
logistic map. By changing some initial conditions of the system, we observed the
transition from order to chaos exhibited by the map.
\end{abstract}


\section{Introduction}
Nonlinear dynamics is a topic that not only cover all the disciplines, in both
natural and social sciences, but also is now becoming part of introductory level
undergraduate courses in sciences. Searching for affordable, easy to setup, and
reconfigurable classroom demonstrations that allow to investigate the physical
and mathematical nature of nonlinear dynamical systems has been always a matter
of interest for instructors. This paper describes a simple apparatus used to
explore the behavior of one-dimensional chaotic maps, like the Logistic Map,
when different initial conditions are chosen.

The device consists of an inexpensive open-source microcontroller connected to
an array of light emitting diodes (LEDs), and programmed to iterate the logistic
map in the one-dimensional interval $[0,1]$. This interval is divided in ten
equal parts and mapped one-to-one to a single LED in the array. When the
logistic map produces certain value after an iteration, the corresponding LED
lights up showing the value in the one dimensional domain approximately. After
several iterations, it is possible to \textit{visualize} the trajectory of the
map by looking at the blinking LEDs sequence. Sensitivity to initial conditions,
density of periodic orbits, strange attractors, and bifurcations are visualized
easily with this device.

\section{One dimensional maps}
The logistic map is, perhaps, the simplest example of how a nonlinear dynamical
equation can give rise to very complex, chaotic behavior.~\citealp{Weisstein}
Initially introduced as a mathematical model of population
growth,~\citealp{Verhulst} it rapidly found applications in diverse areas like
mathematical biology, biometry, demography, condense matter, econophysics, and
computation.~\citealp{Ausloos} The logistic map function is defined as
\begin{equation}\label{Eq:LogisticMap}
  X_{n+1} = A\,X_n (1 - X_n) \equiv f_A(X),
\end{equation}
where the factor $A$ is a model-dependent parameter representing
\textit{external} conditions to the system, and $X_n$ is the population in the
$n$th-period cycle, scaled so that its value fits in the interval
$[0,1]$.~\citealp{Marion} The function $f_A$ is called an \textit{iteration
function} because we may find the population fraction $X$ in the following
period cycle by repeating the mathematical operations expressed in
\eqref{Eq:LogisticMap}.~\citealp{Hilborn} It is also one-dimensional since there
is only a single variable $X$, and the resulting curve is a line.

\section{Setup}
\begin{figure}[h]
\centering
\includegraphics[scale=0.9]{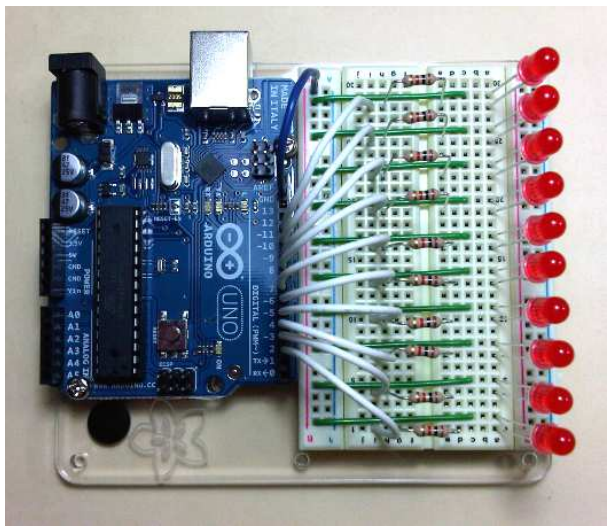}
\caption{\label{Fig01}Setup of the Logistic Map device. An array of ten LEDs are
connected to the Arduino microcontroller using the same number of $220\,\Omega$
resistors.}
\end{figure}

\begin{figure}[h]
\centering
\includegraphics[scale=0.23]{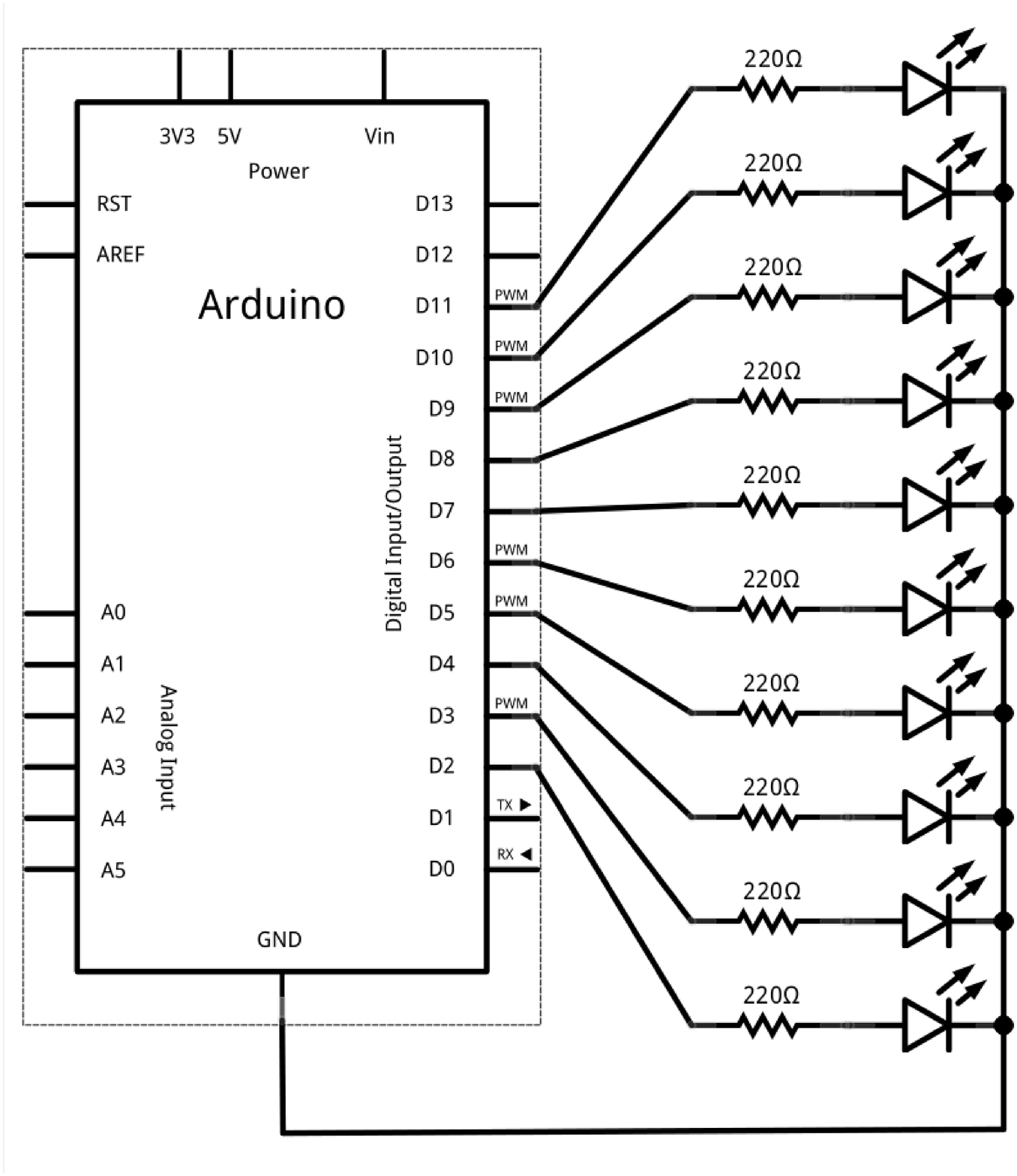}
\caption{\label{Fig02}Circuit diagram of the ten LEDs array connected to the
Arduino.}
\end{figure}

We show a photograph of the apparatus setup in Fig.~\ref{Fig01}, and its
schematics in Fig.~\ref{Fig02}.

The apparatus uses an open-source Arduino prototyping platform made up of an
Atmel AVR processor (microcontroller). It has 14 digital input/output pins, 6
analog inputs, a 16 MHz crystal oscillator, a USB connection, a power jack, an
ICSP header, and a reset button.~\citealp{Arduino} Arduino is a registered
trademark---only the official boards are named ``Arduino''---so clones usually
have names ending with ``duino.''~\citealp{Schmidt} The Arduino can be connected
to a computer through the USB port and programmed using a language similar to
C++. The program is uploaded into the microcontroller using an Integrated
Development Environment (IDE). In the Arduino world, programs are known as
\textit{sketches}.

We connected an array of ten $5\,\mbox{mm}$ LEDs to the digital pins of the
microcontroller using $220\,\Omega$ resistors (or $1\,\mbox{k}\Omega$ for dimmer
LEDs). We also used a solderless breadboard to hook up all the electric
components to the microcontroller.

\section{The sketch}
Listing~\ref{List01} shows a simple program (sketch) used to operate the
microcontroller, and \textit{visualize} the logistic map by looking at the array
of blinking LEDs. A glowing LED represents an iteration of the one-dimensional
map, and it is linked with a value in the interval $[0,1]$.

Arduino programs require two mandatory functions: \texttt{setup()} and
\texttt{loop()}. Any variable or constant defined outside these two functions is
considered to be \textit{global}. In the \texttt{setup()} function, we tell the
microcontroller that there are 10 LEDs connected to the digital pins and that
they are intended to be turned on and off. In the \texttt{loop()} function, the
logistic map is iterated, and the visualization process takes place as we
observe the LEDs turning on and off, one after another, following the evolution
of the nonlinear system.

Using the \texttt{if()} and \texttt{else} control structures, we divide the
interval $[0,1]$ in ten identical segments and associate an LED to each one of
them. For example, the first LED from the left represents the first interval
segment $[0,0.1)$, the second LED represents the segment $[0.1,0.2)$, and so on.
The last LED represents the final segment $[0.9,1.0]$. When the microcontroller
iterates the logistic map, a value belonging to one of these ten intervals is
returned, and the corresponding LED turns on for 500 milliseconds. This process
is repeated infinitely, so we can observe the \textit{orbit} followed by the
logistic map by watching at the blinking LEDs sequence.

\begin{figure}[h]
\centering
\includegraphics[scale=0.3]{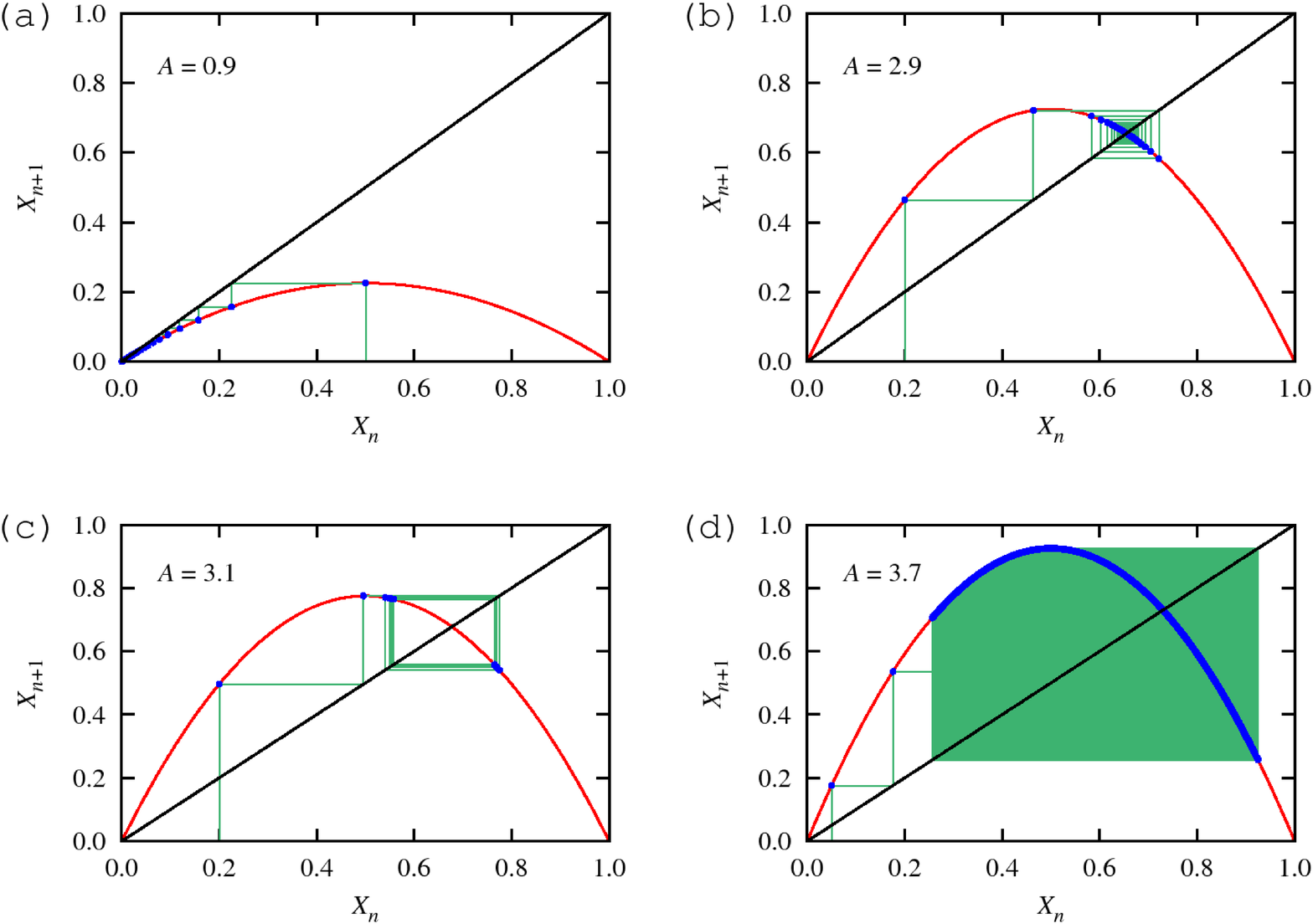}
\caption{\label{Fig03} Cobweb diagrams for the logistic map for different values
of $A$ and $X_0$. The $45^{\circ}$ line, $X_{n+1}=X_n$, is used to find the
fixed points and follow the iteration process. Plots (a) and (b) show the
presence of attractors at $X=0$ and $X=0.655$, respectively. Plot (c) displays a
period doubling bifurcation, and (d) the onset of chaos.}
\end{figure}

The initial conditions for the logistic map can be changed anytime and uploaded
again into the microcontroller. Thus, we can explore the behavior of the chaotic
map when different values of the parameters $A$ and $X_0$ are chosen. It is
amusing to study the bifurcations of the logistic map by looking at the blinking
sequence of LEDs. We start with the parameter values $A=0.9$ and $X_0=0.5$
($0<A<1$). The LED associated with the interval $[0.5,0.6)$ turns on first. As
the iterations of the logistic map take place, we observe that the sequence of
blinking LEDs moves toward the very first LED, which is associated with the
segment $[0,0.1)$. The sequence ends up with this first LED left on
indefinitely. This evolution is shown in Fig.~\ref{Fig03}(a), the cobweb diagram
of the logistic map for that selection of parameters. For this case, $X=0$ is an
\textit{attractor}, and the domain $[0,1]$ forms a \textit{basin of attraction}.

Another example of an attractor happens when $A=2.9$ and $X_0=0.2$ ($1<A<3$). We
see that, after few iterations and blinking LEDs, there is only one LED that
remains on indefinitely. This LED corresponds to the interval $[0.6,0.7)$. In
this case, we have a \textit{fixed point} occurring at $X=0.655$. As it is shown
in Fig.~\ref{Fig03}(b), the orbit of the map follows a square spiral that
converges into that fixed point. Now, if $A=3.1$ and $X_0=0.2$ ($3<A<3.44948$),
we can observe a 2-\textit{period cycle}; the value of $X$ oscillates between
0.558 and 0.765, and the two corresponding LEDs turn on and off intermittently.
Here, the fixed point $X=0.677$ becomes a \textit{repeller}.
Figure~\ref{Fig03}(c) shows the periodic behavior of the logistic map for this
set of parameters. Unfortunately, higher periodicity (like 4-period, 8-period
cycles, or higher) cannot be observed clearly with this device because of the
way the LEDs are mapped into the $[0,1]$ interval. More LEDs are required to
improve the ``resolution'' of the device. However, the microcontroller has a
limit of 13 digital pins, and other electric components (like an extra
microchip) need to be incorporated into the circuit if we want more LEDs
controlled by the device.

Finally, when $A=3.7$ and $X_0=0.05$ ($3.56994<A<4$), chaos
onsets.\citealp{Sprott} An ``apparently'' randomness of blinking LEDs is
observed. After many iterations, there is no single LED that remains on
indefinitely, and no single LED that has not been turned on at least once. The
orbit of the logistic map covers all the domain of the interval $[0,1]$.
Figure~\ref{Fig03}(d) shows all points visited by the logistic map orbit after
$10\,000$ iterations.

\lstset{language=C++,
        basicstyle=\scriptsize,
        commentstyle=\normalfont,
        frame=single,
        captionpos=b,
        morekeywords={pinMode,digitalWrite,delay}
}
\begin{lstlisting}[float=h,caption={Logistic Map sketch for the Arduino.},
  numbers=none,label={List01}]
  // Blinking Logistic Map

  // choose the pin for each LED
  const int NbrLEDs  = 10;
  const int LEDpin[] = {2,3,4,5,6,7,8,9,10,11};

  const int wait = 500;  // wait for 500 milliseconds

  // Logistic Map parameters
  const double A  = 3.7;  // Logistic map constant

  double X0 = 0.2;  // Initial position (0 < X0 < 1.0)
  double X  = X0;   // Use X0 as your first calculated point

  // setup() initializes the LED pins
  void setup() {
      for (int i = 0; i < NbrLEDs; i++) {
          pinMode(LEDpin[i], OUTPUT);
      }
  }

  // loop() iterates the Logistic Map and turn on/off LEDs
  void loop()
  {
      if (X < 0.1)
          blinkLED(LEDpin[0]);
      else if ((X >= 0.1) && (X < 0.2))
          blinkLED(LEDpin[1]);
      else if ((X >= 0.2) && (X < 0.3))
          blinkLED(LEDpin[2]);
      else if ((X >= 0.3) && (X < 0.4))
          blinkLED(LEDpin[3]);
      else if ((X >= 0.4) && (X < 0.5))
          blinkLED(LEDpin[4]);
      else if ((X >= 0.5) && (X < 0.6))
          blinkLED(LEDpin[5]);
      else if ((X >= 0.6) && (X < 0.7))
          blinkLED(LEDpin[6]);
      else if ((X >= 0.7) && (X < 0.8))
          blinkLED(LEDpin[7]);
      else if ((X >= 0.8) && (X < 0.9))
          blinkLED(LEDpin[8]);
      else
          blinkLED(LEDpin[9]);

      // iterates the Logistic Map function
      X0 = X;
      X  = A * X0 * (1.0 - X0);
  }

  // blinkLED function
  // turn on/off LEDs
  void blinkLED(const int pin) {
      digitalWrite(pin, HIGH);  // turn LED on
      delay(wait);              // wait 500 milliseconds
      digitalWrite(pin, LOW);   // turn LED off
  }
\end{lstlisting}

\section{Final comments}
With this apparatus, students may understand more easily the behavior of
one-dimensional chaotic maps looking at an \textit{actual} one-dimensional array
of LEDs; every point in the interval [0,1] produces another point within the
same interval after the logistic map is iterated. By controlling the parameter
$A$ and the initial value $X_0$, the instructor and students can observe under
what conditions periodic and chaotic behavior may occur, providing a better
understanding of periodicity, bifurcations, and the route to chaos of a
nonlinear dynamical system. The setup described here is inexpensive, easy, and
fun to assemble, also enhances computer programming and electronics assembly
skills.


\end{document}